\documentclass[aps,apl,amsfonts,amssymb,superscriptaddress,twocolumn]{revtex4-1}

\usepackage{graphicx}
\usepackage{amsmath}
\usepackage{amssymb}
\usepackage{bm}
\usepackage{color}

\begin{document}

\title{Temperature dependence of Raman-active optical phonons in Bi$_2$Se$_3$ and Sb$_2$Te$_3$ }

\author{Y. Kim}
\affiliation{National High Magnetic Field Laboratory, Tallahassee, FL 32310}
\affiliation{Department of Physics, Florida State University, Tallahassee, FL 32306}

\author{X. Chen}
\affiliation{School of Physics, Georgia Institute of Technology, Atlanta, GA 30332}

\author{Z. Wang}
\affiliation{Department of Physics and Astronomy, University of California, Riverside, CA 92521}

\author{J. Shi}
\affiliation{Department of Physics and Astronomy, University of California, Riverside, CA 92521}

\author{I. Miotkowski}
\affiliation{Department of Physics, Purdue University, West Lafayette, IN 47907}

\author{Y. P. Chen}
\affiliation{Department of Physics, Purdue University, West Lafayette, IN 47907}

\author{P. A. Sharma}
\affiliation{Sandia National Laboratories, Albuquerque, NM 87185}

\author{A. L. Lima Sharma}
\affiliation{Sandia National Laboratories, Albuquerque, NM 87185}

\author{M. A. Hekmaty}
\affiliation{Sandia National Laboratories, Livermore, CA 94550}

\author{Z. Jiang}
\email{zhigang.jiang@physics.gatech.edu}
\affiliation{School of Physics, Georgia Institute of Technology, Atlanta, GA 30332}

\author{D. Smirnov}
\email{smirnov@magnet.fsu.edu}
\affiliation{National High Magnetic Field Laboratory, Tallahassee, FL 32310}

\begin{abstract}
Inelastic light scattering spectra of Bi$_2$Se$_3$ and Sb$_2$Te$_3$ single crystals have been measured over the temperature range from 5 K to 300 K. The temperature dependence of dominant $A^{2}_{1g}$ phonons shows similar behavior in both materials. The temperature dependence of the peak position and linewidth is analyzed considering the anharmonic decay of optical phonons and the material thermal expansion. This work suggests that Raman spectroscopy can be used for thermometry in Bi$_2$Se$_3$- and Sb$_2$Te$_3$-based devices in a wide temperature range.
\end{abstract}

\maketitle

Recently, much attention has been paid to the study of Bi$_2$Se$_3$, Sb$_2$Te$_3$ and other layered stoichiometric compounds, as a promising playground for the realization of a new class of quantum matter, topological insulators \cite{hasan10}. A topological insulator has a ``conventional'' energy gap in the bulk and gapless Dirac-like states on the surface, which are protected against any time-invariant perturbations such as crystal imperfections \cite {hasan09, Zhang09, Hsieh09}. These topologically protected surface states hold great promise for a broad range of potential applications, including field effect transistors \cite{lu_FET_2010,herrero_FET_2010,ong_FET_2010}, infrared-THz detectors \cite{zhangPRB_2010,armitage_2011}, and magnetic field sensors \cite{zhangPRB_2008,vanderbiltPRL_2009}. It is essential to understand the dynamics of phonons in these materials, particularly the phonon-phonon and electron-phonon interactions, in order to achieve the best device performance. While the room-temperature Raman characterizations of optical phonons in Bi$_2$Se$_3$ and Sb$_2$Te$_3$ have been well documented in the literature \cite{Richter77, Richter82}, accurate measurements of the temperature dependence are still lacking. In this Letter, we present a Raman spectroscopy study of Bi$_2$Se$_3$ and Sb$_2$Te$_3$ crystals in the temperature range between 5 K and 300 K. We uncover a characteristic temperature dependence of the phonon peak position and linewidth, and interpret it in the context of thermal expansion and three-phonon anharmonic decay. The observed linear  dependence in the elevated temperature range (80-300 K) allows for applications using Raman spectroscopy as thermometry in Bi$_2$Se$_3$- and Sb$_2$Te$_3$-based devices.

The Bi$_2$Se$_3$ and Sb$_2$Te$_3$ single crystals studied in this work were synthesized at the University of California - Riverside, Purdue University, and the Sandia National Laboratories. Electronic transport characterization revealed that the as-grown crystals were naturally doped, with bulk carrier density in the range of $10^{18}$-$10^{19}$ cm$^{-3}$. The typical onset magnetic field for observing the Shubnikov-de Haas oscillations was found to be $\sim$8 T, corresponding to carrier mobility $\sim$1200 cm$^2$V$^{-1}$s$^{-1}$. 
Raman scattering spectra were collected from fresh crystal surfaces prepared by mechanical exfoliation.

The temperature dependent Raman spectra were measured in a backscattering geometry using a 532 nm laser excitation. The laser light was injected into an optical fiber, guiding the excitation to the sample stage inserted into a helium-flow variable temperature cryostat. The excitation spot size was about 20 $\mu$m in diameter. The scattered light collected by a f/0.73 lens was directed into a collection fiber, and then guided to a spectrometer equipped with a liquid-nitrogen-cooled CCD camera. The spectra were acquired in the spectral region from 110 to 1000 cm$^{-1}$ with a spectral resolution of $\sim$1 cm$^{-1}$. The collected spectra were normalized to the background signal from a mirror to compensate for the effects of parasitic light scattering in the optical fibers. The Raman shift frequency was calibrated with reference to the spectra measured on an elemental sulfur reference sample. 
The peak widths were obtained after correcting for instrumental broadening following the procedure in Ref.\cite{Tanabe81}. To ensure accurate temperature dependent measurements, the temperature was stabilized for approximately 20 minutes at each measurement point before acquiring a spectrum. At a few selected temperatures spectra were recorded during both cooling down and warming up, and were found to be essentially identical within the experimental uncertainty. 

\begin{figure}
\includegraphics[width=8.5cm]{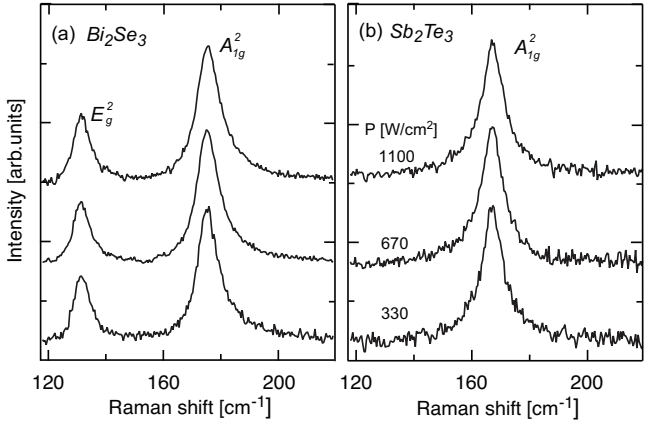}
\caption{\label{fig:300K} 300 K, background corrected Raman spectra of Bi$_2$Se$_3$ and Sb$_2$Te$_3$ measured at different excitation power levels. The spectra are shifted vertically for clarity.}
\end{figure}

Bi$_2$Se$_3$ and Sb$_2$Te$_3$, like other V$_2$VI$_3$ materials, have 5 atoms in a rhombohedral unit cell, and crystalize in the $R\bar{3}m$ ($D^5_{3d}$) structure. Out of 12 optical phonons four modes are Raman active, $2E_{g} + 2A_{1g}$, with the frequencies in the $\sim$30-200 cm$^{-1}$ range \cite{Richter77, Richter82}. Figure 1 shows the room-temperature Raman spectra acquired at different excitation power levels. In Bi$_2$Se$_3$, high frequency $E^{2}_{g}$ and $A^{2}_{1g}$ modes are detected at 131.5 cm$^{-1}$ and 175.5 cm$^{-1}$, respectively. In Sb$_2$Te$_3$, the $E^{2}_{g}$ phonon frequency lies at the low-frequency onset of our apparatus, thus only the $A^{2}_{1g}$ mode at 168.8 cm$^{-1}$ is detected. Great care was taken to ensure that the crystal surface was not affected by local overheating. Up to $\sim$10$^3$ W/cm$^2$, Raman peaks are insensitive to the power level. All the temperature dependent spectra presented below were obtained at power densities of 670 W/cm$^2$ or less.

Typical  Raman spectra measured at fixed temperatures between 300 K and 5 K are shown in Fig.2.   The temperature dependence of the phonon spectra is very similar in both materials.  As the temperature decreases, the phonon peak exhibits blue shift and line narrowing.   In the following, we focus on the temperature dependence of the dominant $A^{2}_{1g}$ phonons. The extracted Lorentzian peak position and FWHM of the $A^{2}_{1g}$ phonon peak are displayed as a function of temperature in Fig.3. 

\begin{figure}
\includegraphics[width=8.5cm]{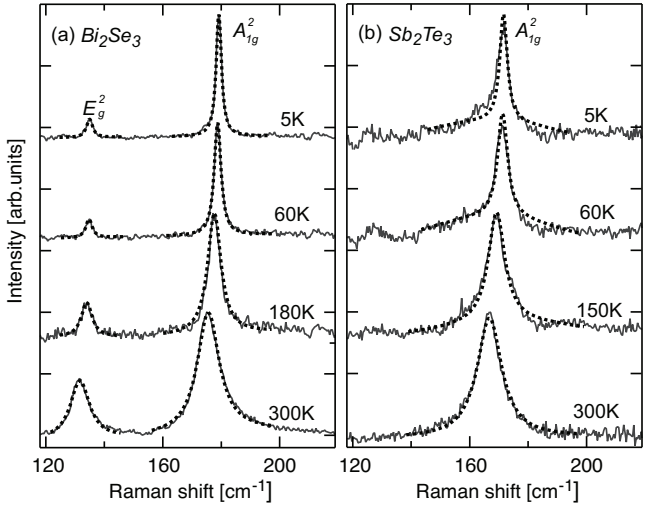}
\caption{ Background corrected, normalized Raman spectra of Bi$_2$Se$_3$ and Sb$_2$Te$_3$ measured at various temperatures. Dashed lines are Lorentzian fits to the experimental data. The spectra are shifted vertically for clarity.}
\end{figure}

In the harmonic approximation, the bare phonon frequency $\omega_{0}$ is obtained from the 2nd order expansion of the lattice potential in normal coordinates. The phonon-phonon coupling leads to the renormalization of the phonon energy and lifetime, which can be described by higher order anharmonic terms assuming the decay of zone-center optical phonon into two or more acoustic phonons. The simplest anharmonic approximation, known as the symmetrical three-phonon coupling model \cite{Klem66}, takes into account the optical phonon decay into two phonons with equal energies and opposite momenta. While it provides a fairly accurate description of the phonon temperature dependence in diamond \cite{diamond00}, more general models have been suggested for other materials accounting for anharmonic contributions due to the thermal expansion and/or asymmetric decay into two or more different phonons \cite{Card84, Balkan83,Tang91}. The temperature dependence of the phonon frequency $\omega(T)$ is commonly expressed as 
\begin{equation}
\omega(T) = \omega_{0} +\Delta\omega^{(1)}(T)+\Delta\omega^{(2)}(T),
\end{equation}
where $\omega_{0}$ is the bare harmonic frequency, $\Delta\omega^{(1)}$ is the anharmonic, sometimes called quasi-harmonic,  correction solely due to the thermal expansion of the crystal lattice, and $\Delta\omega^{(2)}$ is the anharmonic phonon-phonon coupling term. The thermal expansion contribution $\Delta\omega^{(1)}(T)$ is given by 
\begin{equation}
\Delta\omega^{(1)}(T) = \omega_0 \left[\text{exp} \left(-\gamma\int_0^T \left[\alpha_c(T')+2\alpha_a(T') \right] dT' \right) -1\right],
\end{equation}
where $\gamma$ is the mode Gr\"{u}neisen parameter, and $\alpha_a(T)$ and $\alpha_c(T)$ are the coefficients of linear thermal expansion along the $a$ and $c$ axes: $\alpha_L=\dfrac{1}{L}\dfrac{dL}{dT}$.

\begin{figure}
\includegraphics[width=8.5cm]{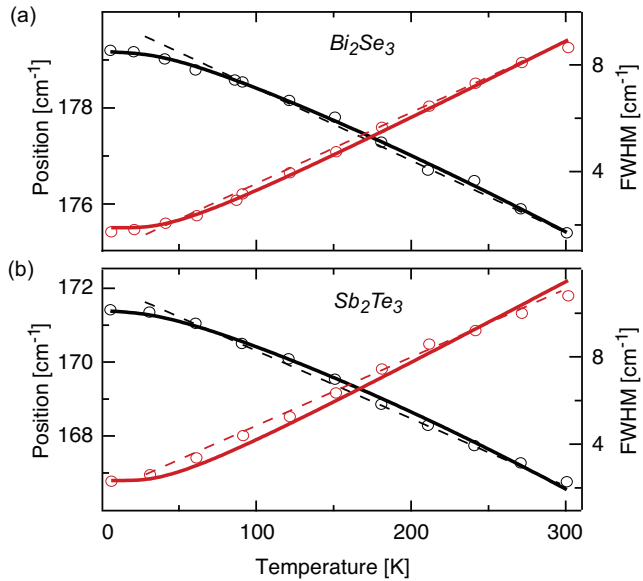}
\caption{Temperature dependence of the $A^{2}_{1g}$ phonon in Bi$_2$Se$_3$. Symbols are peak position (black) and linewidth (red) extracted from the experimental data. The solid lines are fits to the symmetrical three-phonon coupling model including thermal expansion contribution. Dashed lines are linear fits to data in the 90-300 K range.}
\end{figure}

\begin{table*}[ht!] 
\caption{\label{table}Temperature dependence of the $A^{2}_{1g}$ phonon: fitting parameters using the linear model (left) and the symmetrical three-phonon coupling model (right, Eq. 3, 4).}
\begin{ruledtabular}
\begin{tabular}{lcccccccc}
&			\multicolumn{4}{c}{Linear fit (90-300 K)}&								&	\multicolumn{3}{c}{3-phonon decay model} \\
&			$\omega_{0}^*$&	 $d\omega/dT$&		$\gamma_{0}^*$ &	 $d\gamma/dT $&	&	$\omega_{0}$&	A&			B\\
&			(cm$^{-1}$)&	 (cm$^{-1}$/$K$)&		(cm$^{-1}$)&		(cm$^{-1}$/$K$)&	&	(cm$^{-1}$)&	(cm$^{-1}$)&	(cm$^{-1}$)\\
\hline
Bi$_2$Se$_3$&	179.9(6)&		-0.015&			0.8&				0.026&		&	179.8(8)&		-0.7&			1.89 \\
Sb$_2$Te$_3$&	172.2(1)&		-0.018&			1.62&				0.031&		&	172.1&		-0.7&			2.33\\
\end{tabular}
\end{ruledtabular}
\end{table*}

We are not aware of temperature-dependent  data on the thermal expansion coefficients in Bi$_2$Se$_3$ and Sb$_2$Te$_3$. We employed X-ray powder diffraction (XRD) to measure lattice constants of Bi$_2$Se$_3$ and Sb$_2$Te$_3$ at temperatures between 10 K and 270 K. Details about the XRD results and their analysis are described elsewhere \cite{XRD11}. Using the XRD data, we find that  the temperature dependence of lattice parameters in Bi$_2$Se$_3$ can be approximated between 10 K and 270 K with better than 0.8\% accuracy by the following polinomials: 
$a(T)$ = 4.13 + 1.59$\times$10$^{-5}$$T$ + 8.87$\times$10$^{-8}$$T^{2}$, and 
$c(T)$ = 28.48 + 2.26$\times$10$^{-4}$$T$ + 6.86$\times$10$^{-7}$$T^{2}$.
For Sb$_2$Te$_3$, the polynomial approximation with better than 1.5\% accuracy in the same temperature range is given by: 
$a(T)$ = 4.24 + 9.3$\times$10$^{-6}$$T$ + 1.25$\times$10$^{-7}$$T^{2}$, and 
$c(T)$ = 30.18 + 3.5$\times$10$^{-4}$$T$ + 7.4$\times$10$^{-7}$$T^{2}$. 
Estimated   Gr\"{u}neisen parameters  at 270 K are 1.4 for Bi$_2$Se$_3$, and 2.3 for Sb$_2$Te$_3$ \cite{XRD11}. 

We find that the thermal expansion contribution term $\Delta\omega^{(1)}(T)$ accounts for $\sim$40\% of the total phonon frequency change with temperature. Detailed analysis further reveals that the observed temperature dependence can be well described within a symmetrical three-phonon coupling approximation, where the optical phonon frequency $\omega(T)$ and linewidth $\Gamma(T)$ are modeled as
\begin{equation}
\omega(T) = \omega_{0} + \Delta\omega^{(1)}(T) + A\left[ 1 + n(\omega_1) + n(\omega_2) \right],
\end{equation}
\begin{equation}
\Gamma(T) = B\left[ 1 + n(\omega_1) + n(\omega_2) \right],
\end{equation} 
\newline
$\omega_1 = \omega_2 = \omega_0/2$, $n(\omega) = \left[\text{exp} \left( \hbar\omega/ k_BT \right)-1 \right]^{-1}$, and $A$ and $B$ are fitting parameters.  Notice that the linewidth is expressed in terms of anharmonic coupling  only, and does not depend on the thermal expansion contribution. The comparison between the calculated  and the experimental position and linewidth  of the $A^{2}_{1g}$ phonons is shown in Fig. 3. 

Finally, we observe that at  temperatures above 90 K the frequency and linewidth of the $A^{2}_{1g}$ phonon change linearly with temperature, 
$ \omega(T) = \omega_{0}^* + \dfrac{d\omega}{dT} T$,
$ \gamma(T) = \gamma_{0}^* + \dfrac{d\gamma}{dT} T$. 
The overall peak frequency change is about 4 cm$^{-1}$ in the temperature range between 90 K and 300 K, and thus offers an effective way to measure the local temperature. The metrology based on the linewidth measurements could be more accurate because of the larger relative variation, though it would require the use of a high-resolution spectrometer.  The extracted values of the linear fit and the symmetrical three-phonon coupling model are summarized in Table I. 

In conclusion, we performed temperature-dependent Raman spectroscopy measurements on Bi$_2$Se$_3$ and Sb$_2$Te$_3$ crystals. The evolution of the $A^{2}_{1g}$-phonon Raman spectra upon temperature variation is well accounted for by the anharmonic decay of optical phonons and the material thermal expansion.
Above 90 K, the temperature dependence of  the $A^{2}_{1g}$ -phonon Raman peak frequency and linewidth can be very well approximated by a linear function, thus offering a convenient metrology tool  for accurate thermometry in Bi$_2$Se$_3$ and Sb$_2$Te$_3$-based devices in a wide temperature range.

This work is supported by the DOE (DE-FG02-07ER46451). D.S. acknowledges support from the FSU Research Foundation and the NHMFL UCGP-5068.  
Materials synthesis and characterization at UCR and PU are supported by DOE (DE-FG02-07ER46351) and DARPA MESO program, respectively.
The measurements were carried out at the National High Magnetic Field Laboratory, which is supported by NSF Cooperative Agreement No. DMR-0654118, by the State of Florida, and by the DOE.


\begin{thebibliography}{text}

\bibitem{hasan10} For a review, see M. Z. Hasan and C. L. Kane, Rev. Mod. Phys. {\bf 82}, 3045 (2010); X.-L. Qi and S.-C. Zhang, arXiv:1008.2026.

\bibitem{hasan09}Y. Xia, D. Qian, D. Hsieh, L. Wray, A. Pal, H. Lin, A. Bansil, D. Grauer, Y. S. Hor, R. J. Cava, and M. Z. Hasan, Nature Physics {\bf 5}, 398 (2009).

\bibitem{Zhang09}H. Zhang, C.-X. Liu, X.-L. Qi, X. Dai, Z. Fang, and S.-C. Zhang, Nature Physics {\bf 5}, 438 (2009).

\bibitem{Hsieh09} D. Hsieh, Y. Xia, D. Qian, L. Wray, F. Meier, J. H. Dil, J. Osterwalder, L. Patthey, A. V. Fedorov, H. Lin, A. Bansil, D. Grauer, Y. S. Hor, R. J. Cava, and M. Z. Hasan, Phys. Rev. Lett. {\bf 103}, 146401 (2009).

\bibitem{lu_FET_2010} J. Chen, H. J. Qin, F. Yang, J. Liu, T. Guan, F. M. Qu, G. H. Zhang, J. R. Shi, X. C. Xie, C. L. Yang, K. H. Wu, Y. Q. Li, and L. Lu, Phys. Rev. Lett. {\bf 105}, 176602 (2010).

\bibitem{herrero_FET_2010} H. Steinberg, D. R. Gardner, Y. S. Lee, and P. Jarillo-Herrero, Nano Lett. {\bf 10}, 5032 (2010).

\bibitem{ong_FET_2010} J. G. Checkelsky, Y. S. Hor, R. J. Cava, and N. P. Ong, Phys. Rev. Lett. {\bf 106}, 196801 (2011).

\bibitem{zhangPRB_2010} X. Zhang, J. Wang, and S.-C. Zhang, Phys. Rev. B {\bf 82}, 245107 (2010).

\bibitem{armitage_2011} R. Vald\'{e}s Aguilar, A. V. Stier, W. Liu, L. S. Bilbro, D. K. George, N. Bansal, J. Cerne, A. G. Markelz, S. Oh, and N. P. Armitage, arXiv:1105.0237.

\bibitem{zhangPRB_2008} X. L. Qi, T. L. Hughes, and S.-C. Zhang, Phys. Rev. B {\bf 78}, 195424 (2008).

\bibitem{vanderbiltPRL_2009} A. M. Essin, J. E. Moore, and D. Vanderbilt, Phys. Rev. Lett. {\bf 103}, 259902 (2009).

\bibitem{Richter77} W. Richter, H. K\"{o}hler, and C. R. Becker, Physica Status Solidi B-Basic Research {\bf 84}, 619 (1977).

\bibitem{Richter82} W. Richter, A. Krost, U. Nowak, and E. Anastassakis, Z. Phys. B - Condensed Matter {\bf 49}, 191 (1982).

\bibitem{Tanabe81} K. Tanabe and J. Hiraishi,  Appl. Spectrosc. {\bf35}, 436 (1981). 

\bibitem{Klem66} P. G. Klemens, Phys. Rev. {\bf 148}, 845 (1966).

\bibitem{diamond00} M. S. Liu, L. A. Bursill, S. Prawer, and R. Beserman, Phys. Rev. \textbf{B} {\bf 61}, 3391 (2000).

\bibitem{Card84} J. Menendez and M. Cardona, Phys. Rev. \textbf{B} {\bf 29} 2051 (1984).

\bibitem{Balkan83} M. Balkanski, R. F. Wallis, and E. Haro, Phys. Rev. \textbf{B} {\bf 28}, 1928 (1983).

\bibitem{Tang91} H. Tang and I.P. Herman, Phys. Rev. \textbf{B} {\bf 43}, 2299 (1991).

\bibitem{XRD11} X. Chen, H. D. Zhou, A. Kiswandhi, I. Miotkowski, Y. P. Chen, P. A. Sharma, A. L. Lima Sharma, M. A. Hekmaty, D. Smirnov, and Z. Jiang, Appl. Phys. Lett.  (to be published).

\bibitem{Pav11} L. M. Pavlova, Yu. I. Shtern, and R. E. Mironov, High Temperature. {\bf 49}, 369 (2011).

\bibitem{Barnes74} J.O. Barnes, J.A. Rayne, and R.W. Ure, Phys. Lett. A, {\bf 46}, 317 (1974).



\end{thebibliography}
\end{document}